\documentclass[12pt]{article}

\usepackage[export]{adjustbox}
\usepackage{scicite}
\usepackage{setspace}
\usepackage{graphicx}
\usepackage{times}
\usepackage{url}

\newcommand{\beginsupplement}{%
        \setcounter{table}{0}
        \renewcommand{\thetable}{S\arabic{table}}%
        \setcounter{figure}{0}
        \renewcommand{\thefigure}{S\arabic{figure}}%
     }

\newcounter{lastnote}

\title{A machine learning methodology for real-time forecasting of the 2019-2020 COVID-19 outbreak using Internet searches, news alerts, and estimates from mechanistic models}

\author
{Dianbo Liu$^{\dagger,1,2}$, Leonardo Clemente$^{\dagger,1,2,3}$, Canelle Poirier$^{\dagger,1,2}$,\\
Xiyu Ding$^{1,4}$, Matteo Chinazzi$^{5}$, Jessica T Davis$^{5}$, Alessandro Vespignani$^{5,6}$, \\
Mauricio Santillana$^{*,1,2,4}$\\
\\
\normalsize{$^{1}$Computational Health Informatics Program, Boston Children's Hospital, Boston MA 02215}\\
\normalsize{$^{2}$Department of Pediatrics, Harvard Medical School, Boston MA 02215}\\
\normalsize{$^{3}$Tecnológico de Monterrey, 64849, Monterrey, N.L., Mexico} \\
\normalsize{$^{4}$ Harvard T.H. Chan School of Public Health, Boston MA 02215} \\
\normalsize{$^{5}$ Laboratory for the Modeling of Biological and Socio-technical Systems,}\\ \small{Northeastern University, Boston, MA USA}\\
\normalsize{$^{6}$ ISI Foundation, Turin, Italy}\\
\normalsize{$^{*}$ Corresponding author: Mauricio Santillana, msantill@g.harvard.edu}\\
\normalsize{$^{\dagger}$ These authors contributed equally to this paper.}
}

\date{}

\begin{document}

\singlespacing


\maketitle 

\begin{abstract}

\noindent
We present a timely and novel methodology that combines disease estimates from mechanistic models with digital traces, via interpretable machine-learning methodologies, to reliably forecast COVID-19 activity in Chinese provinces in real-time.
Specifically, our method is able to produce stable and accurate forecasts 2 days ahead of current time, and uses as inputs (a) official health reports from Chinese Center Disease for Control and Prevention (China CDC), (b) COVID-19-related internet search activity from Baidu, (c) news media activity reported by Media Cloud, and (d) daily forecasts of COVID-19 activity from GLEAM, an agent-based mechanistic model. Our machine-learning methodology uses a clustering technique that enables the exploitation of geo-spatial synchronicities of COVID-19 activity across Chinese provinces, and a data augmentation technique to deal with the small number of historical disease activity observations, characteristic of emerging outbreaks. Our model's predictive power outperforms a collection of baseline models in 27 out of the 32 Chinese provinces, and could be easily extended to other geographies currently affected by the COVID-19 outbreak to help decision makers.


\end{abstract}

\section*{Introduction}


\noindent
 First detected in Wuhan, China, in December 2019, the SARS-CoV-2 virus had rapidly spread by late January 2020 to all Chinese provinces and many other countries \cite{li2020early,zhu2020novel,chan2020familial,gilbert2020preparedness}. On January 30, 2020, the World Health Organization (WHO) issued a Public Health Emergency of International Concern (PHEIC) \cite{whowebsite,du2020risk,sun2020early,wang2020novel}; and on March 11th, the WHO declared a pandemic for the Coronavirus Disease (COVID-19)\cite{whowebsite}.  By April 6, 2020, the virus had affected more than 1,200,000 people and caused the deaths of 70,000 in more than 180 countries.\cite{sun2020early}.\\
 \\
\noindent
In the last decade, methods that leverage data from Internet-based data sources and data from traditional surveillance systems have emerged as a complementary alternative to provide near real-time disease activity (e.g. Influenza, dengue) estimates \cite{yang2015accurate,santillana2015combining,lu2019improved,cleaton2016characterizing, li2020retrospective}. 
Despite the fact that these methodologies have successfully addressed delays in the availability of health reports as well as case count data quality issues, developing predictive models for an emerging disease outbreak such as COVID-19 is an even more challenging task \cite{lipsitch_enhancing_2019}. There are multiple reasons for this, for example, the availability of epidemiological information for this disease is scarce (there is no historical precedent about the behavior of the disease), the daily/weekly epidemiological reports that become available are frequently revised and corrected retrospectively to account for mistakes in data collection and reporting (a common practice in public health reports), and the presence of a diverse array of uncertainties about disease burden due in part to underreporting of cases \cite{li2020substantial}. \\

\noindent Most efforts to estimate the time evolution of COVID-19 spread and the effect of public health interventions have relied on mechanistic models that parameterize transmission and epidemiological characteristics to produce forecasts of disease
activity\cite{chinazzi2020effect, lai2020effect}. In contrast, few studies have investigated ways to track COVID-19 activity, leveraging internet search data  \cite{li2020early, li2020retrospective, jung2020real}, and none to the best of our knowledge have combined Internet-based data sources and mechanistic estimates to forecast COVID-19 activity. 
\\
\\
\noindent
\textbf{Our contribution}: We present a novel methodology that combines mechanistic and machine-learning methodologies to successfully forecast COVID-19 in real-time at the province level in China. Augmented ARGONet uses a data-driven approach to incorporate inputs from (a) official health reports from China CDC, (b) COVID-19-related internet search activity from Baidu, (c) news media activity reported by Media Cloud, and (d) daily forecasts of COVID-19 activity from GLEAM, an agent-based mechanistic model \cite{chinazzi2020effect}. Inspired by a methodology previously used to successfully forecast seasonal influenza in the United States at the state level \cite{lu2019improved} and previous methods to monitor emerging outbreaks \cite{aiken2019real,mcgough2017forecasting}, Augmented ARGONet is capable of reliably forecasting COVID-19 activity even when limited historical disease activity observations are available. From a methodological perspective, the novelty in our approach comes from a clustering technique that enables the exploitation of geo-spatial synchronicities of COVID-19's activity across Chinese provinces, and a data augmentation technique to mitigate the scarcity of historical data for model-training.


\section*{Results}
\begin{figure}[h!]
    \centering
    \includegraphics[width=0.8\textwidth]{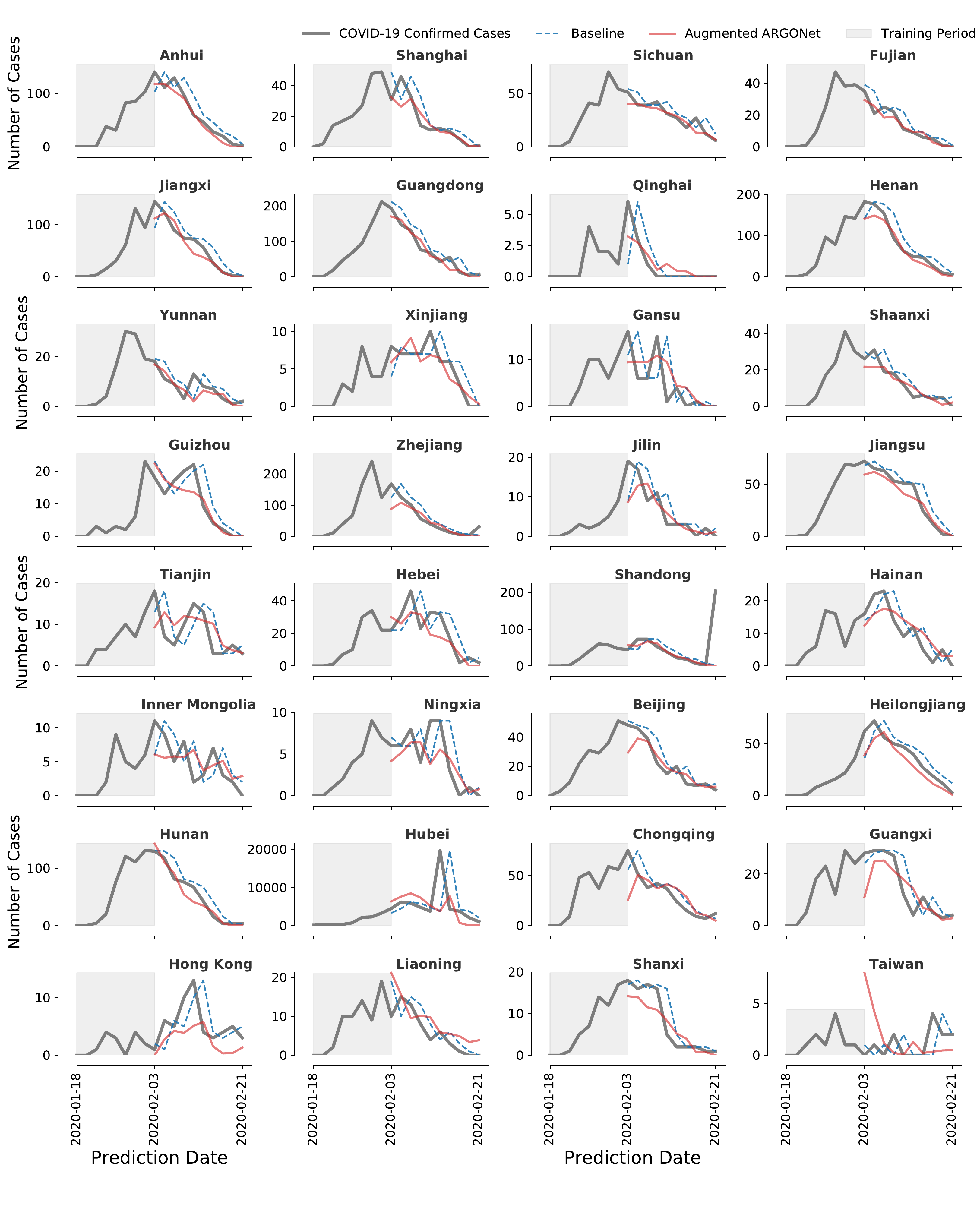}
    \caption{Graphical visualization of the number of new confirmed cases for COVID-19, as reported by China CDC, along with Augmented ARGONet (solid red) 2-day ahead of time estimates between February 3 2020, through February 21 2020. As a comparison, the dotted blue line represents the persistence model.}
    \label{fig:predictions}
\end{figure}

We produced 2-day ahead (strictly out-of-sample) and real-time COVID-19 forecasts for 32 Chinese provinces, for the time period spanning February 3, 2020 to February 21, 2020. In other words, our study was not retrospective, it was conducted as the outbreak unfolded.  The forecasting (case aggregation) time horizon of our predictions was chosen to be 2 days to enhance the signal to noise ratio. Based on geo-spatial synchronicities in COVID-19 activity in each province during the (dynamically increasing) training period, groups of Chinese provinces were clustered, every two days, to train a separate machine learning model per cluster to produce predictions. COVID-19 historical information from official health reports, Baidu search engine frequencies, news reports from Media Cloud, and mechanistic model estimates were then included into separate datasets for each cluster of provinces. Additionally, a data augmentation procedure aimed at increasing the size of each cluster's training dataset was implemented to deal with the scarce number of observations during this relatively-short (local) outbreak. A visual representation of our out-of-sample model forecasts is shown in Figure \ref{fig:predictions} along with the subsequently observed COVID-19 cases, as reported by China CDC.\\

\begin{figure}
    \centering
    \includegraphics[width=1.05\textwidth]{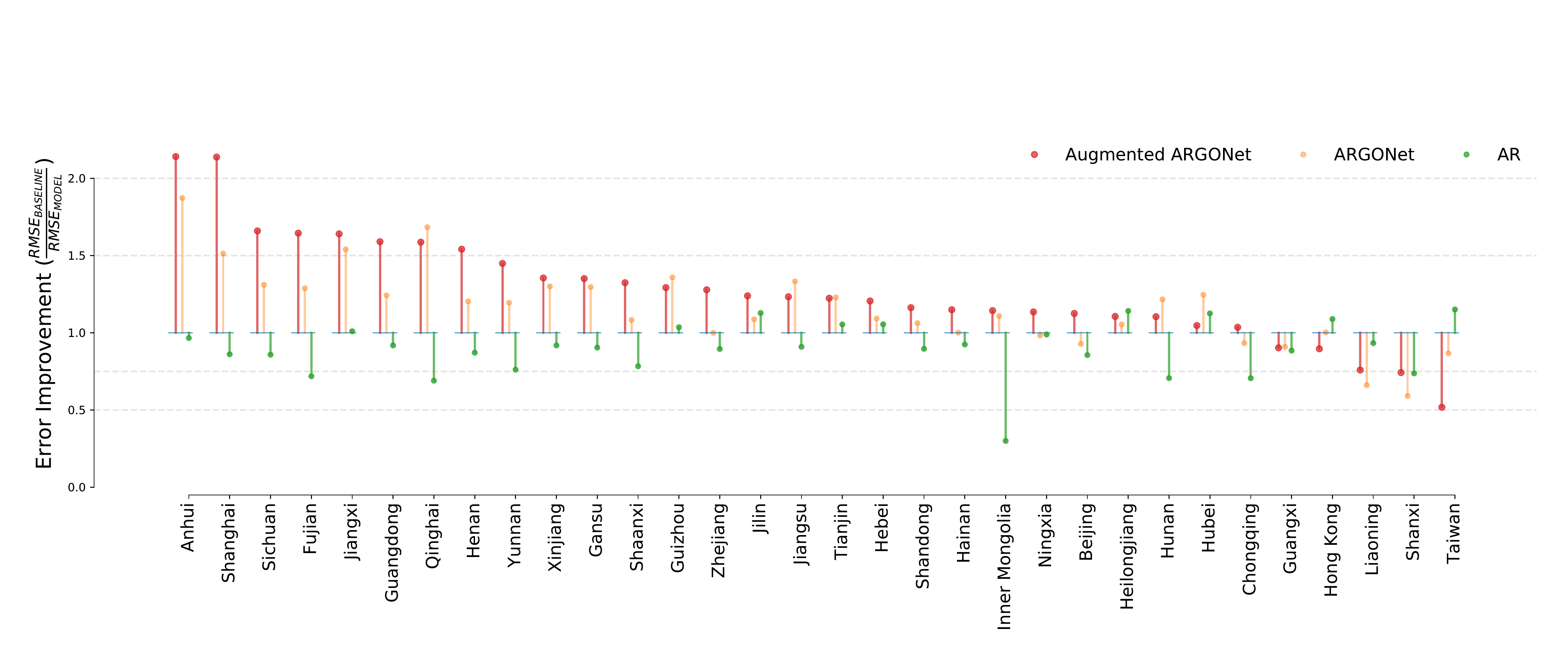}
    \centering
    \includegraphics[width=1.05\textwidth]{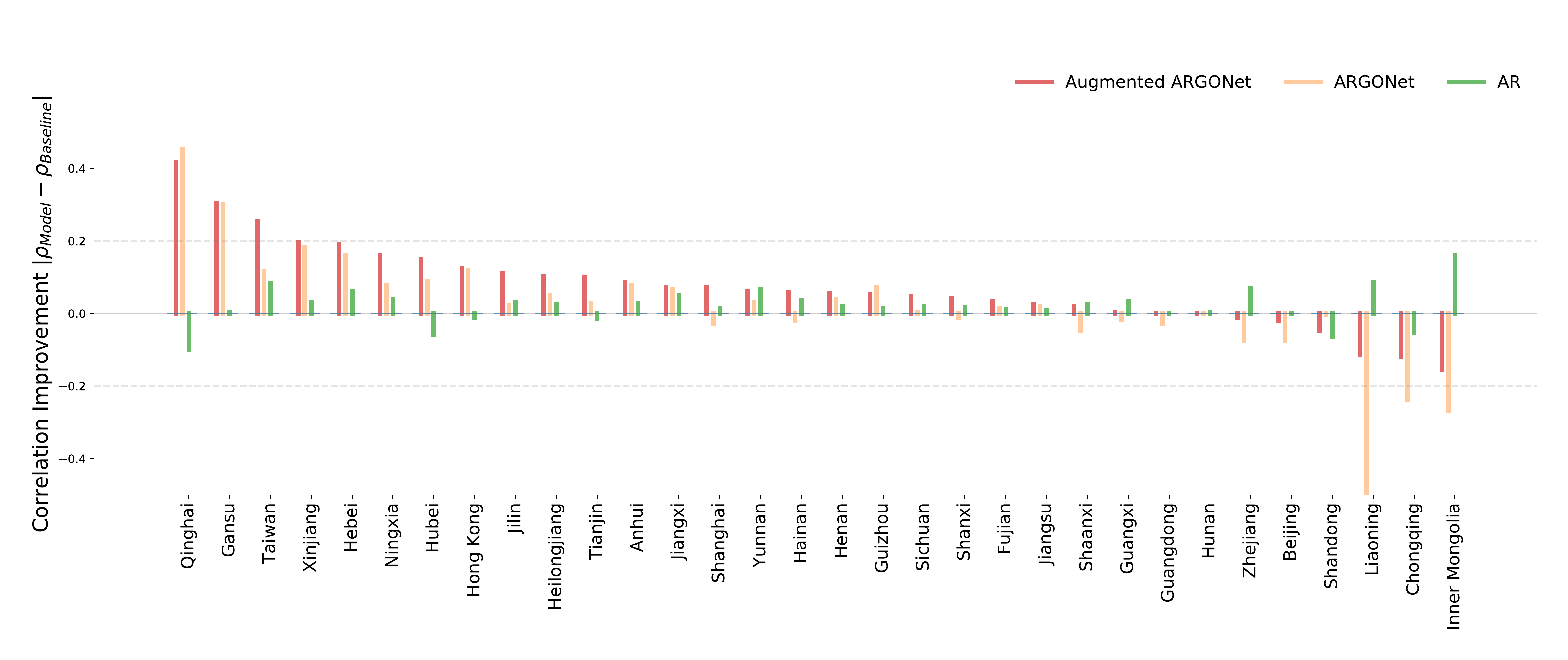}
    \caption{Comparison of the improvement in terms of RMSE (top) and Pearson Correlation (bottom) for each model used in the study. To facilitate comparison between model scores in each province in terms of RMSE, we normalized the RMSE score of each model by the baseline's RMSE and visualized its inverse value. In this way, scores above 1 imply an improvement (RMSE reduction), whereas score below 1 imply the model had a bigger RMSE in comparison to the baseline. In the case of correlation, we plotted the difference between the absolute values between each model's correlation and the baseline. Each panel is ordered, from left to right, based on the metric performance of Augmented ARGONet (solid red).}
    \label{fig:boxplot}
\end{figure}

\noindent To quantitatively evaluate the predictive power of Augmented ARGONet, we compared our forecasts with a collection of baseline forecasts obtained with  (a) a ``persistence model'' that forecasts that the number of new COVID-19 cases (in the next two days) will exactly be the number of cases observed in the past two days (used as the baseline for all models), (b) an autoregressive model that only uses historical COVID-19 activity as input, 
and (c) an enhanced version of ARGONet, without the incorporation of mechanistic model outputs (see details in Materials and Methods section). Our results show that Augmented ARGONet, outperforms the persistence model in 27 out of the 32 Chinese provinces. Even in provinces where Augmented ARGONet failed to produce improvements to the baseline model, our model produced reasonable disease estimates as seen in Figure \ref{fig:predictions}. These provinces include Shanxi, Liaoning, Taiwan, Hong Kong and Guangxi; the latter three with very different administration (and likely healthcare) systems compared to the rest of the provinces.\\
\\
\noindent
\textbf{Models built with only local data do not produce significantly better predictions than the baseline.} 
We analyzed the performance of models built using only local province-level epidemiological data as input. We generated an autoregressive (AR) model for each province, built on COVID-19 cases that occurred in the previous 3 autoregressive lags (i.e. the previous 3, 2-day reports), and compared our estimates with the baseline. Our results, presented in Figure \ref{fig:boxplot} labeled AR, show that the autoregressive model predictive power was overall inferior to the baseline's performance, with exception to Jilin, Tianjin, Hebei, Hubei and Heilongjiang. Subsequently, we incorporated local disease-related internet search information from Baidu and news alert data from media cloud as inputs to build ARGO-type models \cite{yang2015accurate}. These ARGO-type models showed marginal predictive power improvements when compared with autoregressive models (AR) and only outperformed the baseline in 7 provinces.
\\
\\
\noindent
\textbf{Dynamic clustering of Chinese provinces.}  Based on prior work on Influenza activity prediction \cite{lu2019improved}, we added historical COVID-19 activity information for all Chinese provinces to the input of our local models. We calculated the pairwise correlation matrix for confirmed COVID-19 cases between all Chinese provinces, between February 1 and February 21 2020 (Figure \ref{fig:mobility}). Our results showed that most of the provinces experienced similar epidemic trends. 
To build our (clustered) predictive models we combined the data available from several provinces with similar trends (in terms of correlation, which was strictly calculated within our training period at the time-step of prediction). The clustering modeling approach, which incorporated Internet-based data sources as the ARGO-type models,  produced forecasts that led to error reductions for 17 out of 32 provinces compared to the persistence model, and improved correlation values in 20 out of 32 provinces. 
\\
\\
\noindent
 \textbf{Data augmentation.} As an additional way to increase the number of observations in the training set of each cluster, we implemented a data augmentation technique. This process consisted of generating new observations via a Bootstrap method and addition of random Gaussian noise ($\epsilon \sim N(0,0.01)$) to every randomly selected observation. \\
 
 \noindent The results of incorporating both clustering and augmentation techniques can be seen in Figure \ref{fig:boxplot}. For simplicity, we labeled these predictions ARGONet, even though this implementation of ARGONet is an enhanced version specifically designed for emerging outbreaks where data is scarce. In terms of RMSE, our results show that ARGONet's predictive power was able to outperform AR and the persistence model in 25 of the 32 Chinese provinces. In terms of correlation, ARGONet outperformed the baseline (persistence) model in 18 provinces. 
 \\
 \\
 %
 \noindent
 \textbf{Augmented ARGONet.} 
We included forecasts produced by mechanistic model as an additional input in our models (prior to the clustering and augmentation steps). The results of incorporating these estimates can be seen in Figure \ref{fig:boxplot} with the name of Augmented ARGONet.
 Our results show that the inclusion of mechanistic model estimates improved ARGONet's predictive power across the majority of provinces. Augmented ARGONet led to error reductions in 27 out of 32 provinces compared to the baseline. In terms of correlation, it improved in 26 out of 32 provinces. Provinces like Qinghai, Hunan and Jiangxi showed the biggest improvement, whereas Taiwan, Hong Kong, Shanxi and Liaoning did not display error reductions. \\

\begin{figure}
    \centering
    \includegraphics[width=1\textwidth]{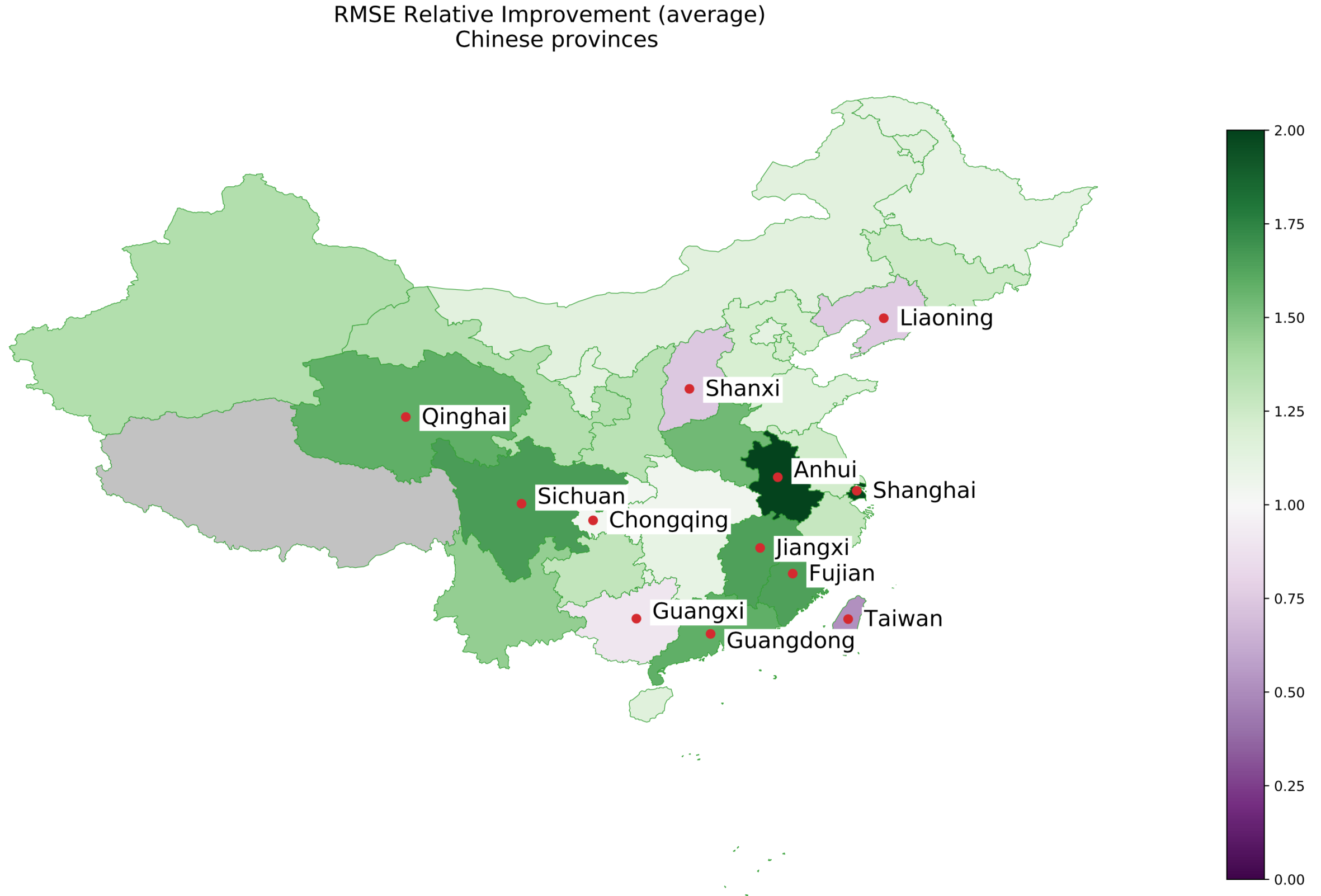}
    \caption{Geographical visualization of the relative improvement ($\frac{RMSE_{model}}{RMSE_{baseline}}$). Chinese provinces that show an increase in performance relative to the baseline are shaded green, while provinces that did not perform better than our baseline are shaded purple. Provinces with the highest improvement (Anhui, Shanghai, Sichuan, Fujian, Jiangxi, Guangdong, Qinghai) and underperformance (Taiwan, Shanxi, Liaoning, Hong Kong, and Guangxi) are identified by a red dot over the province.}
    \label{fig:map}
\end{figure}

\begin{figure*}
    \centering
    \includegraphics[width=1.1\textwidth]{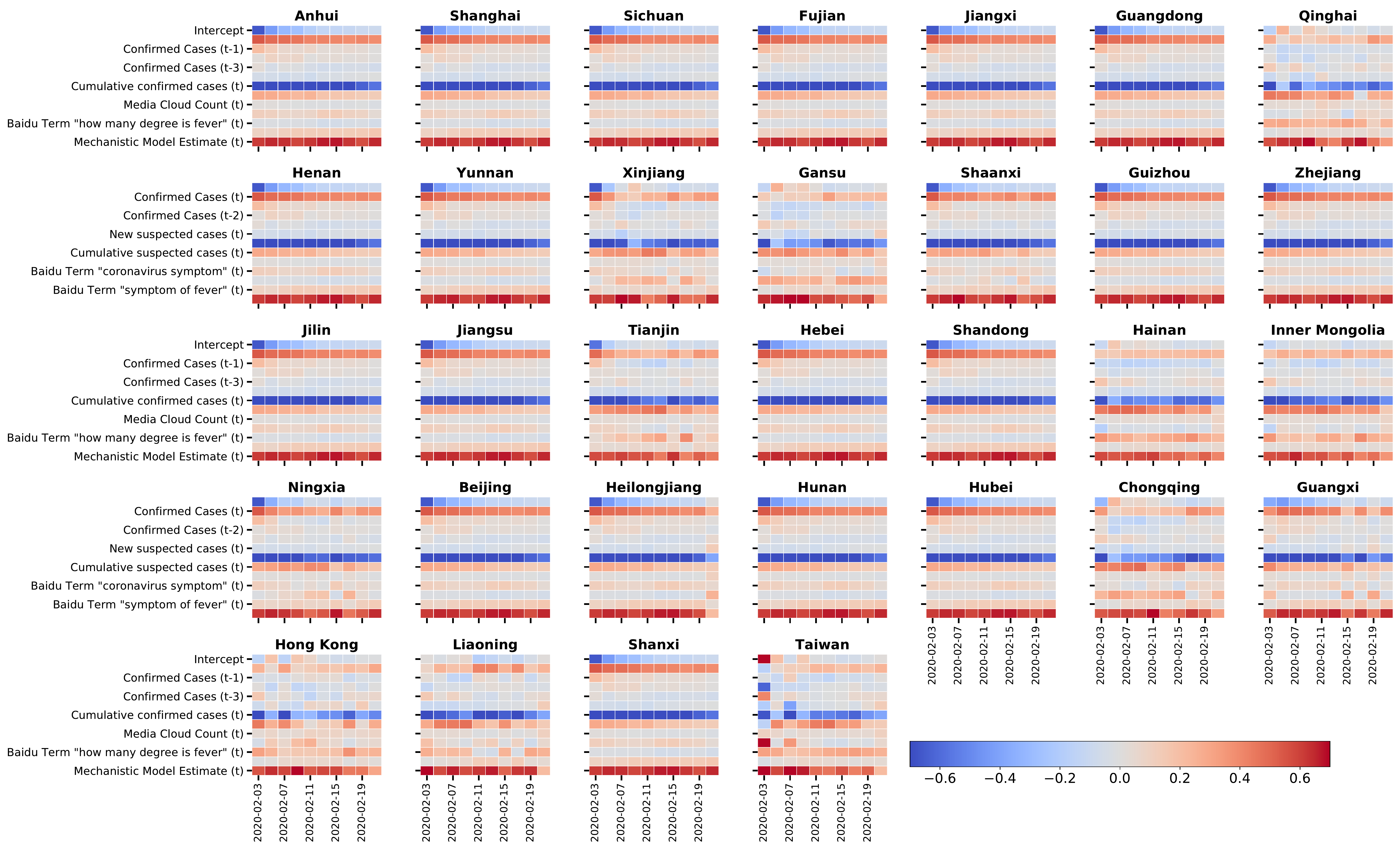}
    \caption{Time evolution of the value (averaged over the 20 experiments) of the linear coefficients for the features used in our methodology, visualized per province. Every heatmap includes the same number of features (rows) and is organized in the same order.}
    \label{fig:heatmap}
\end{figure*}

\noindent As an alternative way to visualize Augmented ARGONet's predictive performance, we plotted a map with Chinese provinces (Figure \ref{fig:map}) color-coded based on the improvement shown in Figure \ref{fig:boxplot}. From a geographical perspective, the provinces where Augmented ARGONet had the most improvement (Anhui, Jiangxi, Fujian, Sichuan, Guangdong) were located in South Central of China. Shanxi, Liaoning, Taiwan, Hong Kong, and Guangxi are the provinces where our models were not able to reduce the error compared to the baseline. While Augmented ARGONet's performance in these provinces was not superior to the baseline, its predictions were within a reasonable range as seen in Figures \ref{fig:predictions} and \ref{fig:boxplot}. We suspect that the reason why performance in Taiwan, Hong Kong, and Guangxi did not improve is due to the different administrative (and likely healthcare) systems of these provinces. We were not able to perform any analysis on Tibet, one of the largest province in China, and Macau given their low count of detected COVID-19 cases.
\\
\\
\noindent
\textbf{Analysis of the importance of the sources of information over time.} To minimize the prediction errors in our estimates, the dynamic design of our methodology utilizes different sources of information as needed over time. This means that, for each province (or group of provinces within a cluster), we can quantify the predictive power of different features used in our models as time evolves. Our analysis, visualized in Figure \ref{fig:heatmap}, shows that historical COVID-19 confirmed cases, and suspected cases were consistently relevant sources of information over most of the study period. Internet-based search terms from Baidu were also frequently used. Daily news counts were used by our models in a selected number of provinces. However, for many of these provinces, the importance of media article counts decrease over time. Estimates from  mechanistic models contributed to our model prediction, especially in early February, 2020.


\clearpage

\section*{Discussion}

We presented a methodology capable of producing meaningful and reliable short-term (2-day ahead) forecasts of COVID-19 activity, at the province level in China, by combining information from reports from China CDC, Internet search trends, news article trends, and information from mechanistic models. Our approach is capable of overcoming multiple challenges characteristic of emerging outbreaks caused by novel pathogens. These challenges include: the lack of historical disease activity information to calibrate models, the low volume of case count data, and the inherent delay to gain access to data. Methodologically speaking,  our method maximizes the use of limited number of observations as the outbreak unfolded by (a) choosing an appropriate aggregation time-window (2 days) to improve the signal to noise ratio , (b) leveraging synchronicities in the spatio-temporal trends in COVID-19, across provinces, to produce cluster-specific models of prediction, and (c) using data augmentation methods to increase the stability in the training of our models.
\\
\\
Previous methods, such as the ARGONet model \cite{lu2019improved, poirier2019influenza}, have been shown to make accurate real-time prediction, at the state-level in the USA for seasonal infectious diseases such as Influenza. In addition, Chinazzi et al.\cite{chinazzi2020effect}, showed that it was possible to estimate the evolution of an emerging outbreak using a mechanistic model. Nevertheless, as far as we know, reliable real-time methodologies to forecast new case counts for an emerging disease outbreak remained an unsolved problem. In this study, we showed that dynamically trained machine learning model can produce accurately real-time estimates for COVID-19 outbreak.
\\
\\
In terms of prediction error, our proposed methodology, Augmented ARGONet, was able to outperform the persistence model in 27 out of 32 provinces. While our method does not show prediction error improvements in Guangxi, Liaoning, Shanxi, Taiwan, and Hong Kong, our forecasts are still within range in all provinces except for Taiwan, where very few cases were reported during the time-period of this study. It is important to note that Taiwan , Hong Kong and Guangxi have different administrative (and likely healthcare) systems compared with other provinces. This, could explain the differences in COVID-19 trends in these regions and could help explain why our models do not seem to add value to the persistence model. Features studies should investigate if incorporating disease activity estimates from other mechanistic models, designed with likely different assumptions and mathematical formulations, could lead to further improvements.\\

\noindent We were unable to identify an accurate (daily) parametrization of changes in human mobility due to the widespread local lockdowns during the period of our study (February 3-21, 2020), and thus, we did not include this data source as a potential predictor. Future studies may incorporate (high temporal resolution) human mobility data as a modulator of transmission and predictor of disease activity. When looking at the entire time-period of this study; however, we observed that the data-driven clustering of provinces used in our approach and based on COVID-19 activity, appears to have similarities with the clustering one would obtain using human mobility data made available by Baidu \cite{mob2020data} (Figure \ref{fig:mobility}). This result aligns with the conclusion of other available studies that find that the time evolution of the COVID-19 outbreak in China, was importantly influenced by changes in human mobility (consequence of public health interventions)\cite{chinazzi2020effect,Kraemer2020mobility,lai2020effect}, and in the early stages, associated to the percentage of people traveling from Wuhan.
\\
\\
Our findings suggest that, it is possible to use very limited amounts of data from multiple data sources to conduct real time forecasting in the early stage of an emerging outbreak. We believe that our method, Augmented ARGONet, could prove to be useful for public health officials to monitor (and perhaps prevent) the spread of the virus\cite{lu2019improved,wang2020novel,wu2020nowcasting,phan2020importation}. As the SARS-CoV-2 virus continues to spread around the world, extensions of our methods could be implemented to provide timely and reliable disease activity estimates to decision makers.

\subsection*{Materials and Methods}

\begin{figure}
    \centering
    \hspace*{-.5cm}
    \includegraphics[width=1\textwidth]{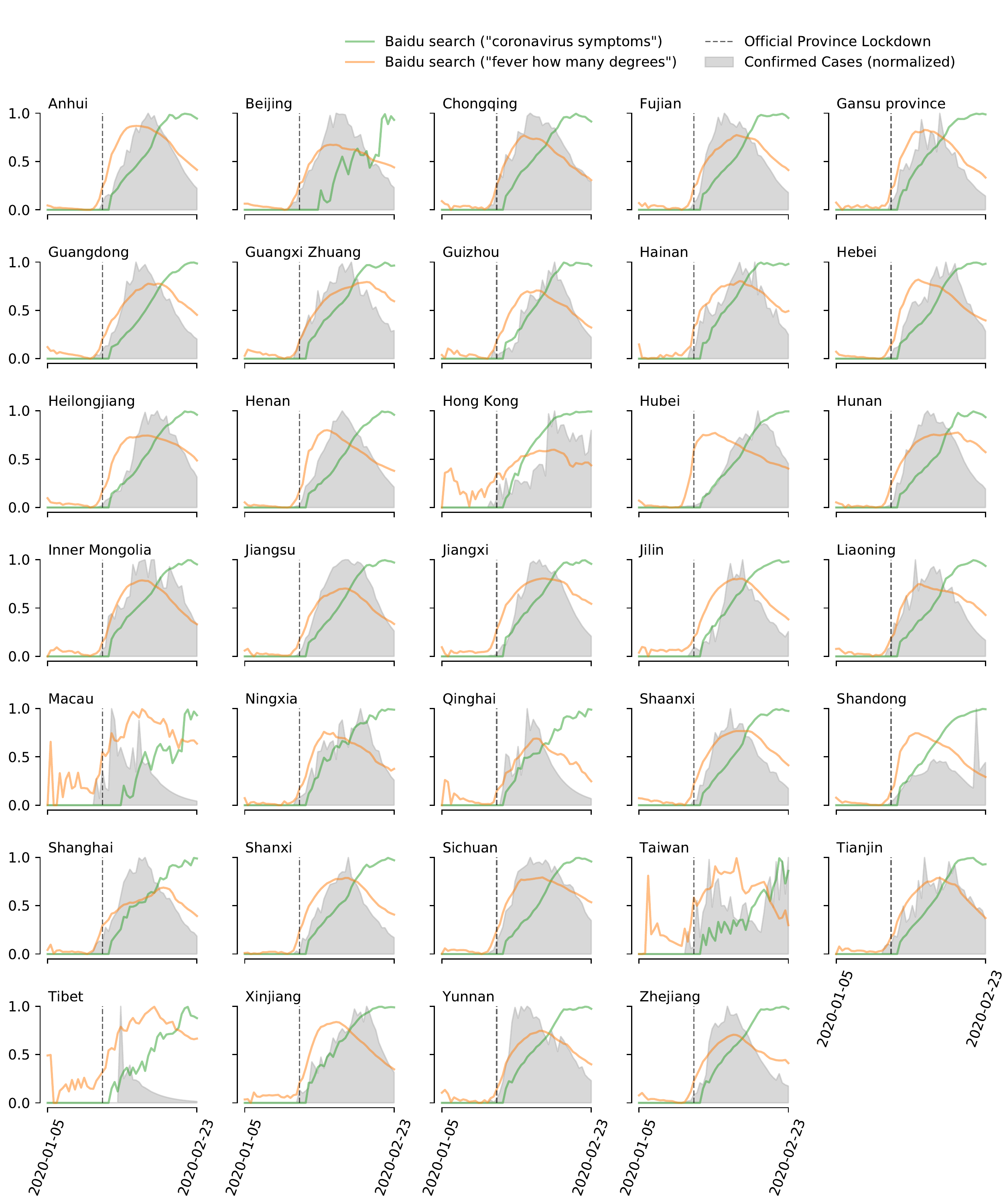}
    \caption{Visualization for the evolution of Covid-19 cases (green and orange), and Baidu search trends (red) in China. All timeseries have been smoothed for visualization purposes.}
    \label{fig:china_baidu_province_timeseries}
\end{figure}

\textbf{Data sources.} We used the following data sources for our study: COVID-19 activity reports from China Centers for Disease Control and Prevention (CDC), Internet search frequencies from Baidu, the number of related news reports from 311 media sources, as reported by the Media Cloud platform, and COVID-19 daily forecasts from the Global Epidemic and Mobility Model. We provide details about each of these data sources in the next paragraphs.
\\
\\
\textbf{Daily reports of COVID-19.} Case counts of COVID-19 were obtained from China CDC . This data is curated and publicly available via the Models of Infectious Disease Agent Study (MIDAS) association (github.com/midas-network/COVID-19/tree/master/data/cases). All data were collected on the original date they became available. Indeed, case counts released by China CDC can be revised, up to several weeks later. In this study, we only used unrevised data, which is the real case scenario to produce real time estimates. 
The reports, available for all the provinces, include various activity trends such as: new diagnosed cases, new suspected cases, and new reported deaths. For our study, we selected the number of confirmed cases as the epidemiological target, and collected activity reports from January 10, 2020 to February 21, 2020. \\
\\
\textbf{Baidu Internet search activity.} We collected the daily search fraction for three different COVID-19-related search terms in Mandarin (translated in English they are "COVID-19 symptoms", "how many degrees" and "symptoms of fever"). These terms were selected based on their correlation and potential association with case counts of COVID-19 \cite{xiaoxuan2020can}, and collected individually for each province from January 1, 2020 to February 21, 2020. Our decision to use Internet activity as a source of information is based on the hypothesis that search frequencies from COVID-19-related keywords reflect, to an extent, the number of people presenting symptoms related to COVID-19 before their arrival to a clinic. Given Baidu imposes limits to the data access for researchers, we were unable to conduct a broad analysis on a wide range keywords. A visualization of the Baidu search term timeseries can be seen in \ref{fig:china_baidu_province_timeseries}. \\
\\
\textbf{News reports}. An online open-source platform called Media Cloud, that allows the tracking and analysis of media for any topic of interest through the matching of keywords, was used. We obtained volumes of the number of news articles available over time from a collection of 311 Chinese media websites using the keywords ``coronavirus" ,``COVID-19", ``2019-nCoV", ``pneumonia",``fever",``cough" and the name of each province to generate province-specific news activity trends. Media data from January 1, 2020 to February 21, 2020 were collected and used as an additional source information. \\

\noindent
\textbf{Aggregation of daily reports}. To enhance signal and reduce noise, we aggregated case count, searches volume and media articles count for each $\delta t=$2 days window. \\
\\
\noindent
As COVID-19 is an emerging outbreak, the amount of epidemiological-related information, either official or non-official, is low, and thus, limits our capacity to build predictive models. To maximize usage of data, we applied the following strategies: \\
\\
\textbf{Clustering.} We clustered the 32 provinces into several groups and train a model for each group. Clustering and model retraining processes were repeated on every single new prediction date.  To know the similarity of outbreak pattern among Chinese provinces, we calculated the pairwise correlation matrix for confirmed COVID-19 cases by using all  historical data available. Then, based on similarity matrix, provinces were clustered by using complete linkage hierarchical clustering, which is an agglomerative hierarchical clustering method, creating clusters based on most dissimilar pairs. \cite{defays1977efficient}. The number of clusters K was determined by choosing the K, maximizing Calinski-Harabasz index\cite{calinski1974dendrite}.\\
\\
\textbf{Data Augmentation.} We conducted data augmentation by using bootstrap method to re-sample  each data point of the training dataset. We made 100 bootstrap samples for each data point to which we added a random Gaussian noise with a mean of 0 and a standard deviation of 0.01.
\\
Due to stochasticity of both clustering algorithm and model training processing, on each prediction day, we run the whole clustering-training process twenty times and take an average of the outputs as our final prediction.\\
\\
\noindent
\textbf{Predictive model}. For our prediction task, we fitted a LASSO multi-variable regularized linear model for every data set generated from our clustering and augmentation steps at time $t$. The LASSO technique minimizes
the mean squared error between observations and predictions subject to a L1 norm constraint. The number of new confirmed COVID-19 cases for the next bi-day can be then expressed as:
\begin{equation}
y_{T+\delta t}=\sum_{i=0}^{3}\alpha_{i}y_{T-i \delta t}+\beta S_{T}+\gamma M_{T}+\delta D_{T} + \psi C_{T} + \epsilon_{T+\delta t} 
\end{equation}

\noindent
where 
\begin{itemize}
\item $y_{T+ \delta t }$ is the estimate at date $T+\delta t $,where $\delta t =2$ days
    \item $y_T$ is the number of cases at date $T$.
    \item $S_{T}$ is the searches volume at date $T$.
    \item $M_{T}$ is the number of media article at date $T$.
    \item $D_{T}$ is the number of deaths at date $T$.
    \item $C_{T}$ is the number of cumulative cases at date $T$.
    \item $\epsilon_{T+\delta t}$ is the normally distributed error term.
\end{itemize}
 Model were dynamically recalibrated, similar to the method presented in Santillana et al. 2014 \cite{santillana2014using} and Lu et al. 2019 \cite{lu2019improved}. Our method was implemented in R 3.5.3  environment with glmnet 3.0-2 library. \\
\\
\textbf{Global Epidemic and Mobility Model (GLEAM) } GLEAM, the mechanistic model used in this study, is an individual-based, stochastic, and spatial epidemic model \cite{chinazzi2020effect,balcan2010modeling,gomes2014pastore,ZhangE4334}. GLEAM uses a meta-population network approach which was integrated with real-world data. In GLEAM, the world population is divided into sub-populations centered around major transportation hubs (usually airports). The sub-populations are connected by the flow of individuals traveling daily among them. Over 3,000 sub-populations in about 200 different countries and territories are included in the model. The airline transportation data consider daily origin-destination traffic flows obtained from the Official Aviation Guide (OAG) and IATA databases (updated in 2019). Ground mobility flows are derived by the analysis and modeling of data collected from the statistics offices for 30 countries on 5 continents~\cite{balcan2010modeling}. Mobility variations in Mainland China were adapted from information from Baidu Location-Based Services (LBS). Within each sub-population, the human-to-human transmission of COVID-19 is modeled using a compartmental representation of the disease where each individual can occupy one of the following four states: Susceptible ($S$), Latent ($L$), Infectious ($I$) and Removed ($R$). Susceptible individuals can acquire the virus through contacts with individuals in the infectious state, and become latent, meaning they are infected but can not transmit the infection yet. Latent individuals progress to the infectious stage with a rate inversely proportional to the latent period. Infectious individuals progress into the removed stage with a rate inversely proportional to the infectious period. The sum of the average latent and infectious periods defines the generation time. Removed individuals represent those who can no longer infect others, meaning they were isolated, hospitalized, died, or have recovered. \\ 
The model produces an ensemble of possible epidemic scenarios described by the number of newly generated infections, times of disease arrival in each subpopulation, and the number of traveling infection carriers. We make an assumption that a starting date of the epidemic that falls between November, 15 2019 and December 1, 2019, with 40 cases IS caused by zoonotic exposure\cite{Rambauts,ferguson-40,Anderson,Bedford}. The transmission dynamic is calibrated by using an Approximate Bayesian Computation approach to estimate the posterior distribution of the basic reproductive number $R_0$\cite{Sunnaker}. We assume that the world has a detection of imported cases as low as 40\%~\cite{Niehus2020.02.13.20022707,Salazar2020.02.04.20020495}. Data on importation of cases were derived from currently available published line lists\cite{Sun2020,Pinotti2020.02.24.20027326}.
\\
\\
\textbf{Performance of model and relevance of predictors.} Two different metrics were used to measure the performance of our predictive model : 1) The root mean square error (RMSE), 2) The Pearson correlation.
To assess the predictive power of our methodology, we compared our performance against the following models:

\begin{enumerate}
    \item Persistence rule (Baseline): A rule-based model that uses the new case count at date $T$ as estimate of prediction for $T+\delta t$ so that ($ {y}_{T+\delta t} = y_T$)
    \item Autoregressive (AR): A simple autoregressive model built on COVID-19 cases that occurred in the previous 3 autoregressive lags (2-day reports). Please refer to  Supplementary Materials for more information on this model.
    \item ARGONet : An alternate version of our methodology that does not include any mechanistic information.
    
\end{enumerate}
As linear models are used in this study, relevance of predictors in predicting new cases can be defined thanks to the associated factor of each term in the trained model. As all data were normalized using the z-score (strictly within the training datasets) during training and prediction, the associated factor can be approximately understood as,  how many standard deviations the predicted new cases $y_{T+\delta t}$ will change if 1 standard deviation change in the predictor.

\section*{Acknowledgments}
We thank Dr. Wei Luo for his assistance and guidance on the interpretation of mobility data for Chinese provinces. MC and AV acknowledge support from Google Cloud Healthcare and Life Sciences Solutions via GCP research credits program.

\section*{Funding statement}
CP, AV, and MS were partially supported by the National Institute of General Medical Sciences of the National Institutes of Health under Award Number R01GM130668. The content is solely the responsibility of the authors and does not necessarily represent the official views of the National Institutes of Health.

\section*{Author contributions}
DL, LC, CP, AV, and MS conceived and designed the study. DL, LC, CP, XD, MC collected the different data sources.
MC, JTD, and AV, produced predictions using the GLEAM modeling platform. DL, LC, CP, implemented the Augmented ARGONet methodology. DL, LC, CP, and MS analyzed the results. DL, LC, CP, and MS wrote the first draft of the manuscript. All authors contributed to and approved the final version of the manuscript.

\section*{Competing interests}

The authors have declared no competing interest.

\section*{Data sharing}

All codes and data will be made available via the Harvard dataverse.


\bibliographystyle{unsrt}
\bibliography{References}

\section*{Supplementary}
\subsection*{Mobility data correlation}
\beginsupplement

\begin{figure}
    \centering
    \includegraphics[width=1\textwidth]{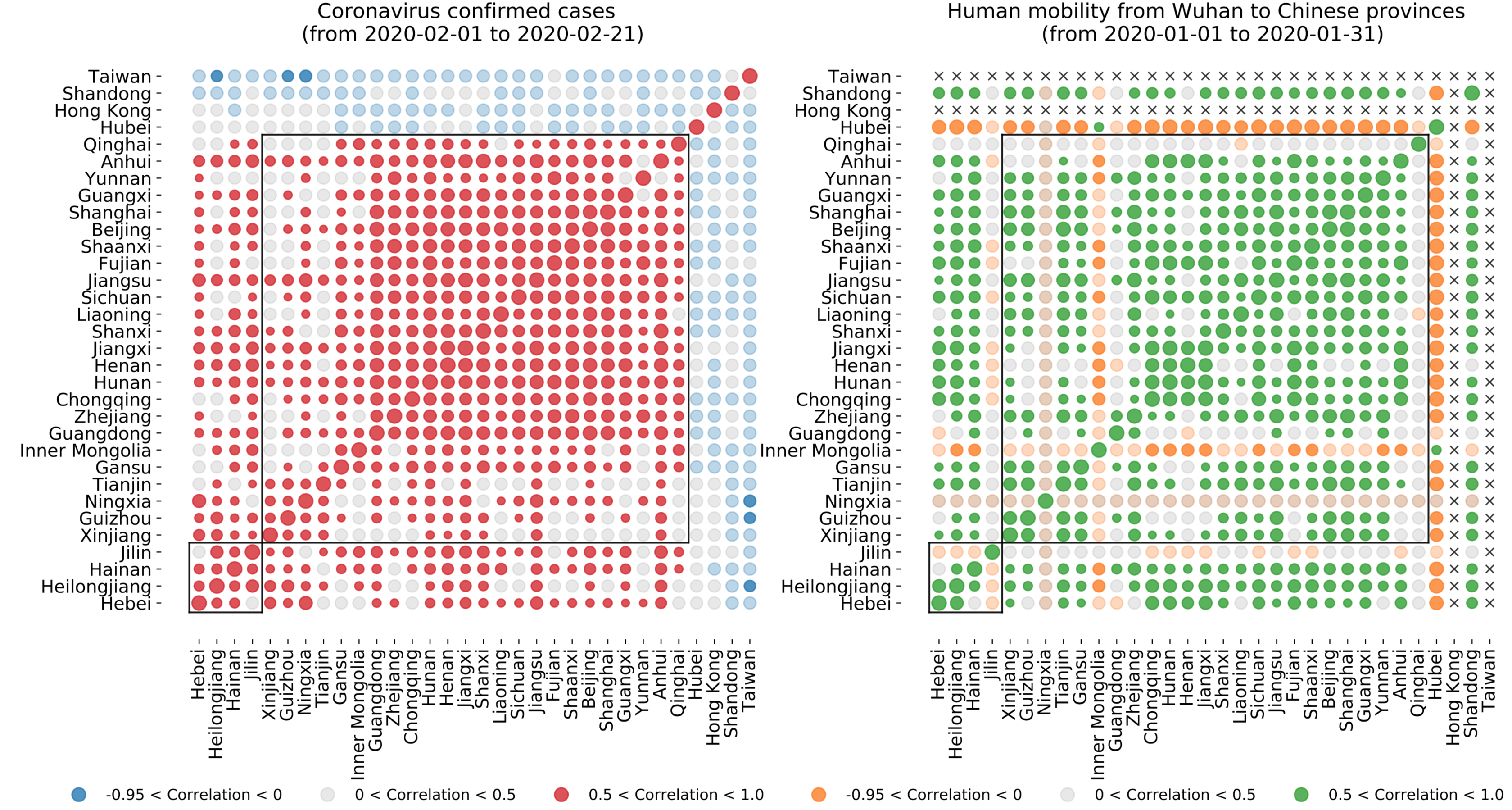}
    \caption{Visualization of the pairwise correlation matrix of human mobility from Wuhan to each Chinese province. During the period of January, we can see a similar trend of mobility for a big cluster of provinces (the mobility matrix has been rearranged in the same province order than the COVID-19 case counts correlation matrix to facilitate visualization of the similarity).}
    \label{fig:mobility}
\end{figure}

\textbf{Human mobility data}. As an additional source, we collected the daily human mobility (in terms of percentage) between the Chinese provinces via Baidu. We decided to not include human mobility information as a source of information for our methodology given the absence of data during February.\\
\\
To investigate the impact of human mobility on new confirmed cases, we performed a similar analysis on Baidu mobility data. Baidu mobility data was obtained from a public available archive \url{(https://doi.org/10.18130/V3/YQLJ5W)} \cite{mob2020data}, which was originally scraped from \url{(qiangxi.baidu.com)} website. The dataset contains daily inbound (i.e. percentage of people traveling to the city from all the cities in China) and outbound (i.e. percentage of people traveling from the city to all the cities in China) mobility data for all cities (except Hong Kong, Macau and Taiwan) in China on each day from January 1, 2020 to January 31, 2020. \\
\\
For this study, we were particularly interested in the relationship between the amount of people traveling from Wuhan/Hubei to other provinces and the number of new confirmed cases in those provinces. Given the original data from Baidu is city-level, we first extracted all the outbound data from Wuhan (which is the capital of Hubei and the first city with confirmed COVID-19 cases) and mapped all the cities belonging to the same province to get province-level outbound mobility from Wuhan. We obtained a 31x31 matrix, where each row is a province in China, and each column is one day in January, each element shows the percentage of people traveling out of Wuhan to the other provinces, among all people traveling from Wuhan on that day.\\
\\
To understand the similarity of the human mobility from Wuhan among Chinese provinces and investigate the potential impact of mobility on the confirmed cases of each province, we calculated the pairwise correlation matrix for the percentage of people traveling from Wuhan to each province by using the data from January 1, 2020 to January 31, 2020 (see Figure \ref{fig:mobility}). If the pairwise correlation between the two provinces is high, it indicates that the trend in the percentage of people traveling from Wuhan to the two provinces are similar. We calculated the pairwise correlation matrix for new confirmed cases using data from February 1, 2020 to February 21, 2020, and we found that provinces which are geographically closed are not necessarily clustered together, indicating geographical distance would not be able to explain the similarity in the new confirmed cases. To further explore the potential impact of mobility on the new confirmed cases of each province, we then calculated the Pearson correlation and cosinus similarity between the two correlation matrices. To do this, we first converted the two matrices into two vectors and calculated the correlation between the two vectors. We got Pearson correlation of 0.1473 and cosinus similarity of 0.4729.

\begin{figure}
    \centering
    \includegraphics[width=1\textwidth]{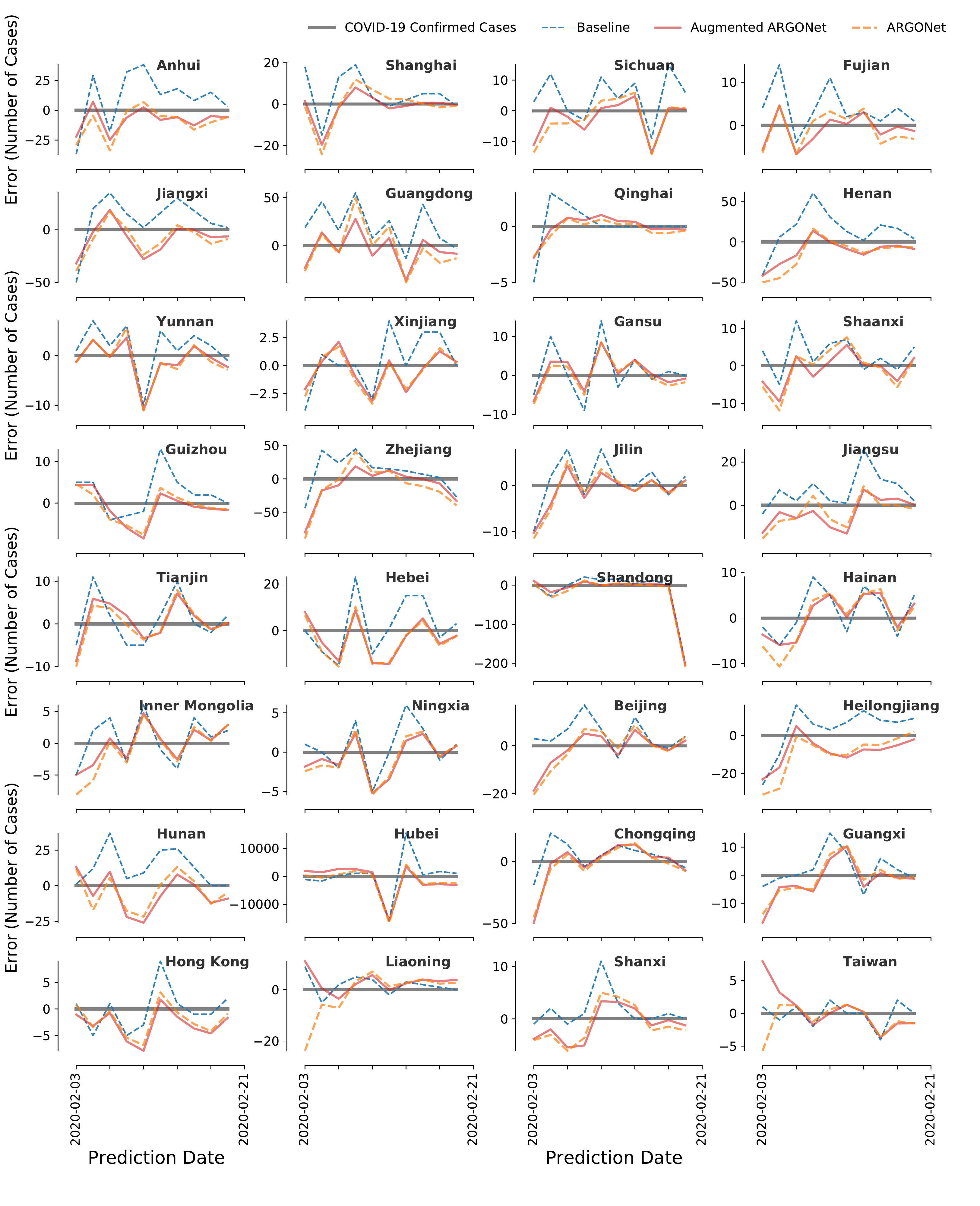}
    \caption{Graphical visualization of the out-of-sample COVID-19 error ($\hat{y} - y$) between February 03, 2020, to February 21, 2020.}
    \label{fig:error}
\end{figure}
\clearpage
\subsection*{Autoregressive Model}
An autoregressive model (AR) was built using only COVID-19 local (in contrast to Augmented ARGONet's clustering methodology) historical confirmed case counts from the past three 2-day reports. Based on this formulation, our 2-day forecast ($\hat{y}_{t+2}$) of COVID-19 confirmed cases is estimated as follows: 
\begin{equation}
 \label{eqejem}
  \begin{array}{c}
y_{T+\delta t}=\sum_{i=0}^{3}\alpha_{i}y_{T-i \delta t} + \epsilon_{T+\delta t}
  \end{array}
\end{equation}
where 
\begin{itemize}
\item $y_{T+ \delta t }$ is the estimate at date $T+\delta t $,where $\delta t =2$ days
    \item $y_T$ is the number of cases at date $T$.
    \item $\epsilon_t \sim  N(0, \sigma^{2})$, residuals
\end{itemize}

As for Augmented ARGONet, our version of AR was dynamically trained.

\end{document}